\documentclass[letterpaper, 10 pt, conference]{ieeeconf}

\IEEEoverridecommandlockouts                              

\usepackage[utf8]{inputenc}
\usepackage{include_packages} 
\usepackage[font=footnotesize,skip=5pt]{caption}
\usepackage[english]{babel}

\title{Collaborative Coalitions in Multi-Agent Systems: Quantifying the Strong Price of Anarchy for Resource Allocation Games}
\author{{Bryce L. Ferguson, Dario Paccagnan, Bary S. R. Pradelski, and Jason R. Marden}
\thanks{This research was supported by ONR Grant \#N00014-20-1-2359, AFOSR Grant \#FA95550-20-1-0054 and \#FA9550-21-1-0203}%
\thanks{B. L. Ferguson (corresponding author) and J. R. Marden are with the Department of Electrical and Computer Engineering, University of California, Santa Barbara, CA, {\texttt{\{blferguson,jrmarden\}@ece.ucsb.edu}}.}%
\thanks{D. Paccagnan is with the Department of Computing, Imperial College London, {\texttt{\{d.paccagnan\}@imperial.ac.uk}}.}%
\thanks{B. S. R. Pradelski  is with the National Center for Scientific Research (CNRS), {\texttt{\{bary.pradelski@cnrs.fr\}@cnrs.fr}}.}%
}

\begin{document}
\maketitle

\begin{abstract}
	The emergence of new communication technologies allows us to expand our understanding of distributed control and consider collaborative decision-making paradigms.
	With collaborative algorithms, certain local decision-making entities (or agents) are enabled to communicate and collaborate on their actions with one another to attain better system behavior.
	By limiting the amount of communication, these algorithms exist somewhere between centralized and fully distributed approaches.
	To understand the possible benefits of this inter-agent collaboration, we model a multi-agent system as a common-interest game in which groups of agents can collaborate on their action to jointly increase the system welfare.
	We specifically consider $k$-strong Nash equilibria as the emergent behavior of these systems and address how well these states approximate the system optimal, formalized by the $k$-strong price of anarchy ratio.
	Our main contributions are in generating tight bounds on the $k$-strong price of anarchy in finite resource allocation games as the solution to a tractable linear program.
    By varying $k$ --the maximum size of a collaborative coalition--we observe exactly how much performance is gained from inter-agent collaboration.
    To investigate further opportunities for improvement, we generate upper bounds on the maximum attainable $k$-strong price of anarchy when the agents' utility function can be designed. 
%
%
%
\end{abstract}

\section{Introduction}
Controlling large-scale systems like transportation services \cite{Wollenstein2020}, robotic fleets \cite{khamis2015multi}, supply chains \cite{ranganathanReInventingFoodSupply2022}, or machine job schedules \cite{Tsakalozos2011} is often challenging as managing each component simultaneously can be difficult if not impossible.
As a method of reducing control algorithm complexity --or to satisfy various physical and technical constraints-- a system operator may opt to distribute decision-making across the system's components~\cite{antonelli2013interconnected}.
Doing so forms a multi-agent system, where each agent makes system-relevant decisions locally.
Distributed control for multi-agent systems has been relevant and highly studied in many domains and proven effective at providing reasonable system behavior~\cite{antonelli2013interconnected,mardenCooperativeControlPotential2009,murray2007recent}.
However, the idea of fully localizing decision-making is perhaps excessive.
New technologies allow for greater communication and coordination between components of large-scale systems~\cite{das2019tarmac,Ferguson2022_Cost}.
These new technologies open the door to new opportunities to control multi-agent systems with \emph{collaborative decision-making algorithms}, in which decision-making is no longer fully distributed but can be partially coordinated between the agents of the system.
How these control paradigms compare to the more well-studied distributed approach is not well understood. 
In this work, we seek to answer the following questions:
%
%
%
%
\begin{enumerate}
\item[(i)] How does collaborative decision-making affect the behavior of a multi-agent system?
\item[(ii)] Can performance be improved by intelligently designing how a coalition makes decisions?
\end{enumerate}
We model the multi-agent system as a common-interest game in which the agents' joint-action leads to some system welfare.
In the fully distributed setting, the price of anarchy serves as a quantitative metric to compare the behavior that emerges from distributed decision-making algorithms to the optimal system performance~\cite{Koutsoupias2009}.
Additionally, by designing the agents utility function, this performance can be improved upon~\cite{Paccagnan2018a}.
We introduce collaborative coalitions to the decision-making problem, where groups of agents can coordinate their actions to jointly maximize the system objective.
These collaborations will alter the behavior of the multi-agent system and ideally lead to improved performance.
Further, we can similarly design the utilities of a group of agents and quantify the additional gains of using optimal collaborative designs.

The manner in which agents communicate and collaborate is context dependent, e.g., networked communication \cite{saad2009coalitional}, pair-wise \cite{bayram2013multirobot}, local \cite{cappello2021distributed}, etc.
To gain general insights on the value of collaboration, we consider that each subset of agents up to size $k$ can collaborate on their action.
Though extensive, this subsumes several other types of communication and can help us understand the possible value inter-agent collaboration can provide.
Further, by varying $k$, we can probe the specific benefit of various scales of connectivity.
In these collaborative environments, 
a stable state of the system is that of the \emph{$k$-strong Nash equilibria}, in which no subset of agents up to size $k$ can revise their actions to receive a higher payoff for each member of the group~\cite{aumann1959acceptable}.
Researchers have studied the existence \cite{nessah2014existence} and computation \cite{gatti2017verification} of strong Nash equilibria in settings including congestion games \cite{holzman1997strong}, lexicographical games \cite{harks2009strong}, and Markov games \cite{clempner2020finding}.
In this work, we study how well \kstrong Nash equilibria approximate the optimal solution, termed the \hbox{\emph{$k$-Strong Price of Anarchy}} ($k$-SPoA).

Quantifying the \kstrong price of anarchy has been studied in network formation games~\cite{epsteinStrongEquilibriumCost2007,andelmanStrongPriceAnarchy2009} and load balancing games \cite{fiat2007strong,epstein2012price,chien2009strong}, as well as more general utility maximizing games~\cite{bachrachStrongPriceAnarchy2014,feldman2015unified}.
In many of these works, the bounds are either not tight or only hold in the limit of large numbers of players.
In this work we devise tight bounds on the \kstrong price of anarchy in games with finitely many players.
Additionally, we study how the strong price of anarchy can be optimized by designing how coalitions of agents make decisions, something previously unstudied in the literature.

%

We specifically focus our analysis on resource allocation games.
In the fully distributed setting ($k=1$), researchers have been successful in quantifying the price of anarchy for Nash equilibria and devising how agents should make decisions~\cite{Gairing2009,Paccagnan2018a}.
However, existing techniques do not hold for collaborative decision-making as actions are no longer made unilaterally and simply characterizing an equilibrium becomes more difficult.
Despite this, we identify several problem reductions that lead to the following results:

\begin{hangparas}{.25in}{1}
\textbf{\cref{thm:quant} (Quantify SPoA):} We generate tight bounds on the \kstrong price of anarchy in resource allocation problems by the value of a tractable linear program.
\end{hangparas}
\begin{hangparas}{.25in}{1}
\textbf{\cref{prop:design} (Optimize SPoA):} We extend this linear program to give an upper bound on \kstrong price of anarchy under the optimal utility design.
\end{hangparas}
\noindent In \cref{fig:spoa}, we plot the \kstrong price of anarchy in maximum coverage problems for $1 \leq k \leq n$ to illustrate the exact benefit collaborative decision-making and design can provide.

\section{Decision-Making Model}\label{sec:model}
Before discussing a specific problem setting, we will review the ideas of distributed decision-making and collaboration in a more general framework for multi-agent systems.
For notation, $x^{\underline{y}} = x!/(x-y)!$ when $x \geq y\geq 0$ and zero otherwise, and $[n] = \{1,\ldots,n\}$.
\subsection{Collaborative Decision-Making}\label{subsec:collab}
Consider a finite set of agents $N = \{1,\ldots,n\}$.
Each agent $i \in N$ selects an action $a_i$ from a finite action set $\mathcal{A}_i$.
When each agent selects an action, we will denote their joint-action by the tuple $a = (a_1,\ldots,a_n) \in \mathcal{A} = \mathcal{A}_1\times\cdots\times\mathcal{A}_n$.
Let $G = (N,\mathcal{A})$ be a tuple encoding the components of a multi-agent system.
The agents' actions ultimately dictate the systems performance, as such, for each joint-action $a$ we assign a system welfare $W(a)$ where $W:\mathcal{A}\rightarrow \mathbb{R}$ is the system designer's objective function.
The system designer would like to configure the agents to reach an action that maximizes the system welfare, i.e.,
\begin{equation}
\Opt{a} \in \argmax_{a \in \mathcal{A}} W(a).
\end{equation}
Though this system state is ideal, it may be difficult to attain as 1) solving for the optimal allocation can be combinatorial and in some cases (including those from \cref{sec:ra}) NP-hard~\cite{Gairing2009}, and 2) it requires a centralized authority to control all agents, which may be practically or logistically difficult.
To resolve this, we will consider that agents make decisions in a decentralized manner.

Fully distributing the decision-making involves designing each agent to locally update their action and has been widely studied and shown effective at providing reasonable system behavior \cite{antonelli2013interconnected}; however, it need not be necessary as emerging communication technologies enable \textit{collaborative inter-agent decision-making}\cite{das2019tarmac}.
To do so, a system operator must make two decisions: 1) which group of agents can collaborate on their decisions (possibly subject to some operational constraints), and 2) how should the agents collaborate on their decisions.
A natural choice for the latter is a group best response.
Let $\Gamma \subseteq N$ be a group of agents endowed with the ability to collaboratively select a \textit{group action} $a_\Gamma \in \mathcal{A}_{\Gamma} = {\prod}_{i \in \Gamma} \mathcal{A}_i$, which they select by maximizing the system welfare over their possible group actions,
\begin{equation}\label{eq:gamma_welf_update}
    a_{\Gamma} \in \argmax_{a_\Gamma^\prime \in \mathcal{A}_\Gamma} W(a_\Gamma^\prime,a_{-\Gamma}),
\end{equation}
where $a_{-\Gamma}$ denotes the actions of the players not in $\Gamma$.
If their are multiple elements in the argmax, the group breaks them at random unless they can remain with their current action.
In \cref{subsec:design}, we will consider other ways the group can make a decision.
Regardless, one would imagine that the greater the collaborative structure, the greater the impact on emergent behavior.

For the system operator's decision over which groups should collaborate, let $\mathcal{C} \subseteq 2^N$ denote the \emph{collaboration set}, or the set of groups of agents ($\Gamma \in \mathcal{C}$) able to collaborate their decisions.
These collaborations can overlap--where agents can partake in multiple, disparate collaborations--and vary in size.
For example, if agents send signals through a communication network~\cite{saad2009coalitional}, we will have \hbox{$\mathcal{C} = \{ (i,j) \in N^2 \mid (i,j) \in E\}$} where $E$ are the edges in a communication graph.
If agents are allowed to communicate with each other one at a time and make pairwise decisions~\cite{bayram2013multirobot}, then $\mathcal{C} = \{ (i,j) \in N^2\}$.
If agents can only communicate with others within a local proximity~\cite{cappello2021distributed}, then $\mathcal{C} = \{ \Gamma \subseteq N \mid \rho(i,j) \leq d~\forall i,j\in \Gamma\}$ where $\rho$ measures the distance between two agents and $d$ is a maximum communication range.
Once the system operator decides on the collaborative structure and the group decision-making protocol, the agents' decision-making process forms a collaborative multi-agent system, denoted by the tuple $(G,W,\mathcal{C})$.
Future work will study the costs of greater collaboration; in this work, we seek to understand what benefit it can provide.
We study a particular choice of collaborative structure where agents collaborate in groups up to size $k$.

\subsection{Strong Nash equilibria}\label{subsec:strong}
Let \hbox{$\mathcal{C}_k = \{\Gamma \subseteq N \mid |\Gamma| = k \}$} denote the subsets of exactly $k$ agents and $\mathcal{C}_{[k]} = \bigcup_{\zeta \in [k]} \mathcal{C}_\zeta$ be the subsets that contain at most $k$ agents.
As $k$ increases, so too does the size and number of collaborating groups.
Varying $k$ between $1$ and $n$ will inform us of the added benefit of increased collaboration.

For some $k$, the states that are stable under collaborative decision-making processes are those
where no $k$-lateral deviation increases the system objective $W$.
These states align with the set of \kstrong Nash equilibria in the common-interest game with payoffs $W$ and player/actions $G$.
\begin{definition}\label{def:ksnash}
A joint-action $\KSNash{a} \in \mathcal{A}$ is a \kstrong Nash equilibrium for the common-interest game $(G,W,\mathcal{C}_{[k]})$ if
\begin{equation}\label{eq:ksnash_def}
    W(\KSNash{a}) \geq W(a^\prime_\Gamma,\KSNash{a}_{-\Gamma}),~\forall a^\prime_\Gamma \in \mathcal{A}_\Gamma,~\Gamma \in \mathcal{C}_{[k]}.
\end{equation}
\end{definition}
Let $\KSNE(G,W) \subseteq \mathcal{A}$ denote the set of all \kstrong Nash equilibria in the game $(G,W,\mathcal{C}_{[k]})$.
When $k=1$, this recovers the definition of a Nash equilibrium that is highly studied in the fully distributed setting~\cite{Koutsoupias2009,Gairing2009,Paccagnan2018a}.
When $k=n$, agents fully collaborate and the only \kstrong Nash equilibria are those that maximize the system welfare.
When $1<k<n$, there is only partial collaboration among the agents, and the resulting system behavior is not yet well understood.
This is in part because agents can participate in multiple collaborations; this overlap is valuable because it helps connect the decisions in different parts of the system (even if not directly) and is why we do not simply partition the agents into disjoint groups.

\cref{def:ksnash} differs slightly from the more general definition in the literature in which groups take actions that are Pareto-optimal over the individual payoffs of the group members~\cite{aumann1959acceptable}.
The definitions are equivalent when considering common-interest games.
This refinement is natural for our setting as we seek to design multi-agent systems which maximize a central objective.
Additionally, in general a \kstrong Nash equilibrium need not exist, however in our setting that is not the case.
\begin{proposition}\label{prop:exist}
In a common-interest game $(G,W,\mathcal{C}_{[k]})$, for any $k \in [n]$, a $k$-strong Nash equilibrium exists and the set $\KSNE$ is stable and reached in finite time by the deterministic coalition best response dynamics and almost surely by the asynchronous coalition best response dynamics.
\end{proposition}
The proof of \cref{prop:exist} is straightforward\footnote{A more general result holds that there exists an equilibrium for any collaboration set $\mathcal{C}$, not just the \kstrong Nash equilibria.}; a more detailed version appears in the appendix.
The claims follow from $\Opt{a}$ being a \kstrong Nash equilibrium for any $k$ and that action revisions strictly increase the welfare, making the set $\KSNE$ absorbing in both the deterministic or asynchronous dynamics.
Certainly, the algorithmic and communication complexity increases with $k$; future work will investigate this cost further.
In this work we seek to understand the possible benefit of inter-agent collaboration.

\cref{def:ksnash} tells us that a \kstrong Nash equilibrium is a local optimum of $W$ in the neighborhood of $k$-lateral deviations.
Still, if $k < n$, this need not be the global optimal.
To quantify the efficacy of \kstrong Nash equilibria as a solution concept, we consider the \textit{\kstrong price of anarchy}
\begin{equation}
    \spoa_k(G,W) = \frac{\min_{\KSNash{a} \in \KSNE(G,W)} W(\KSNash{a})}{\max_{\Opt{a} \in \mathcal{A}} W(\Opt{a}) } \in [0,1].
\end{equation}
In the multi-agent system $(G,W,\mathcal{C}_{[k]})$ any \kstrong Nash equilibrium approximates the optimal solution at least as well as $\spoa_k(G,W)$.
Because $k^\prime {\rm SNE}  \subseteq \KSNE$ for $k^\prime > k$, the \kstrong price of anarchy will be non-decreasing in $k$, i.e., the \kstrong Nash equilibrium will do a better job approximating the optimal solution with greater collaboration.

\subsection{Summary of Contributions}
\begin{figure}
    \centering
    \vspace{1mm}
    \includegraphics[width=0.485\textwidth]{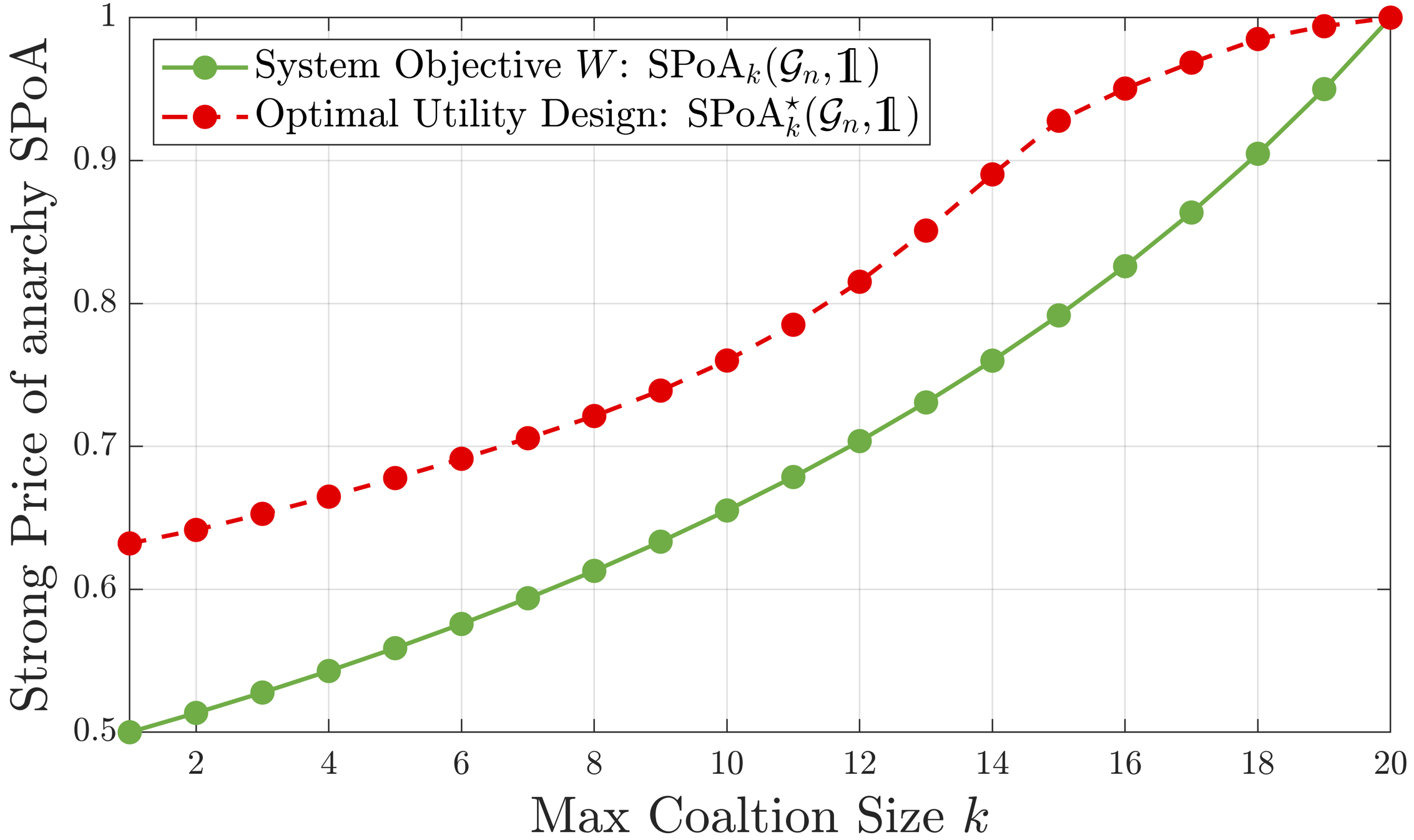}
    \caption{Strong Price of Anarchy in resource allocation games with $n=20$ players and coalitions up to size $k$ (horizontal axis). As the size of groups that are allowed to collaborate grows, so too does the approximation ratio (i.e., strong price of anarchy) of a \kstrong Nash equilibrium. The efficiency of an equilibrium can be further improved by designing the utility functions agents are set to maximize. The solid green line is the \kstrong price of anarchy when agents maximize the system objective (generated by \cref{thm:quant}). The dashed red line is an upper bound on the \kstrong price of anarchy while using an optimal utility design (generated by \cref{prop:design}.)}
    \label{fig:spoa}
    \vspace{-6mm}
\end{figure}

With the possibility of collaboration, an equilibrium becomes more difficult to characterize than in the fully distributed setting.
We circumvent this by introducing a parameterization which allows us to generalize the $\mathcal{O}\left(\sum_{\zeta = 1}^k{n \choose \zeta} m^\zeta\right)$ comparisons of \eqref{eq:ksnash_def} (where $m := \max_{i \in N} |\mathcal{A}_i|$) into $k$ linear inequalities.
This reduction is key in \cref{thm:quant}, in which we show the tight bound on \kstrong price of anarchy is the solution to a tractable linear program for a broad class of resource allocation problems.
In \cref{fig:spoa}, we plot the bound on the \kstrong price of anarchy in maximum coverage problems for $1 \leq k \leq n$ to quantify the performance gained by collaborative decision-making.

Finally, we investigate whether further improvement is possible with collaborative decision-making by designing how a group of agents makes decisions.
Specifically, we consider that the agents' need not maximize the system welfare as in \eqref{eq:gamma_welf_update}, but could instead maximize a separately designed utility function.
Unlike the fully distributed setting, we may desire that groups of different sizes maximize different objectives.
In \cref{subsec:design}, we show that the aforementioned linear program can be extended to provide an upper bound on the maximum attainable \kstrong price of anarchy by utility design in resource allocation problems.
\cref{fig:spoa} plots these values and demonstrates the possible improvements from intelligently designing the collaborative decision-making.

\begin{figure*}
\begin{align}
P^\star = &\max_{\theta \in \mathbb{R}^{|\mathcal{I}|}_{\geq 0}} \quad && \sum_{e,x,o}w(o+x)\theta(e,x,o)\nonumber\\
&~~{\rm s.t.} && \sum_{e,x,o}\left( \frac{n!}{(n-\zeta)!}w(e+x)- \sum_{\substack{0\leq\alpha\leq \zeta\\ 0\leq\beta\leq \zeta-\alpha}} {\zeta\choose \alpha} {\zeta-\alpha \choose \beta} e^{\underline{\alpha}}o^{\underline{\beta}}(n\hs-\hs e\hs -\hs o)^{\underline{\zeta\text{-}\alpha\text{-}\beta}} w(e\hs +\hs x\hs +\hs \beta\hs -\hs \alpha)  \right)\theta(e,x,o) \geq 0\nonumber\\
& && ~~~~~~~~~~~~~~~~~~~~~~~~~~~~~~~~~~~~~~~~~~~~~~~~~~~~~~~~~~~~~~~~~~~~~~~~~~~~~~~~~~~~~~~~~~~~~~~~~\forall\zeta \in \{1,\ldots,k\}\nonumber\\
& && \sum_{e,x,o}w(e+x)\theta(e,x,o)=1\tag{P}\label{opt:primal}
\end{align}
\vspace{-8mm}
\end{figure*}
\section{Resource Allocation Problems}\label{sec:ra}
\subsection{Definition}\label{subsec:define}
To provide more context to the value of inter-agent collaboration, we will consider the \kstrong price of anarchy in a class of resource allocation problems.
Consider a set of resources or tasks $\mathcal{R} = \{1,\ldots,R\}$, to which agents are assigned, i.e., agent $i \in N$ selects a subset of these resources as its action $a_i \subseteq \mathcal{R}$ from a constrained set of subsets $\mathcal{A}_i \subseteq 2^\mathcal{R}$.
Each resource $r \in \mathcal{R}$ has a value $v_r \geq 0$; the welfare contributed by a resource is $v_r w(|a|_r)$, where $w: \{0,\ldots,n\} \rightarrow \mathbb{R}_{\geq 0}$ captures the added benefit of having multiple agents assigned to the same resources and $|a|_r$ is the number of agents assigned to $r$ in allocation $a$.
Assume that $w(0) = 0$ and $w(y)>0$ for all $y > 0$.
The system welfare is thus
\begin{equation}\label{eq:ra_welf}
    W(a) = \sum_{r \in \mathcal{R}} v_r w(|a|_r).
\end{equation}
The objective of the system operator is thus to find an allocation $\Opt{a}$ that maximizes \eqref{eq:ra_welf}; however, doing so in general is NP hard~\cite{Gairing2009} and requires full coordination of the agents.
To alleviate this, we consider the distributed approach described in \cref{sec:model}.
More specifically, we will consider the \kstrong Nash equilibrium solution concept described in \cref{subsec:strong} and investigate how well these states approximate the optimal solution.

Let $G = (\mathcal{R},N,\mathcal{A},\{v_r\}_{r \in \mathcal{R}})$ denote a resource allocation problem (a more specific multi-agent problem than as in \cref{sec:model}).
To understand how \kstrong Nash equilibria perform in this class of problems let $\gee_n$ denote the set of resource allocation problems with $n$ agents.
The \kstrong price of anarchy for this class of problems is defined as
\begin{equation}\label{eq:spoa_bound}
    \spoa_k(\gee_n,w) = \min_{G \in \gee_n} \spoa_k(G,w).
\end{equation}
This performance ratio is parameterized by our choice of welfare function $w$ and the size of collaborative coalitions $k$.
When $k=1$, we recover the traditional price of anarchy for Nash equilibria.
In \cref{subsec:quant}, we will quantify the \kstrong price of anarchy for resource allocation games.
In \cref{subsec:design}, we investigate what further opportunities are present by designing the agents objective function separately from the system objective to alter the set of equilibria.

\subsection{Strong Price of Anarchy}\label{subsec:quant}
To understand the possible benefit of inter-agent collaboration, in~\cref{thm:quant} we quantify the \kstrong price of anarchy in resource allocation problems with the solution to a tractable linear program.
\begin{theorem}\label{thm:quant}
For the class of resource allocation problems $\gee_n$ with welfare function $w$, when agents maximize the common-interest welfare,
\begin{equation}\label{eq:thm_quant}
    \spoa_k(\gee_n,w) = 1/P^\star(n,w,k),
\end{equation}
where $P^\star(n,w,k)$ is the value of the linear program \eqref{opt:primal}.
\end{theorem}
The proof appears in the appendix.
\eqref{opt:primal} is a linear program with decision variable $\theta \in \mathbb{R}_{\geq 0}^{|\mathcal{I}|}$, where $\mathcal{I} = \{(e,x,o)\in \mathbb{N}_{\geq 0}^3 \mid 1 \leq e+x+o \leq n\}$ serves as an index set with size on the order of $n^3$.
Any sum that is listed over $(e,x,o)$ is assumed to be for each $(e,x,o) \in \mathcal{I}$.
The proof possesses three main steps: 
1) we take the original problem \eqref{eq:spoa_bound} and perform a series of reductions allowing us to write the \kstrong Nash equilibrium condition as $k$ inequalities,
2) a game is parameterized by a vector $\theta \in \mathbb{R}_{\geq 0}^{|\mathcal{I}|}$, turning the \kstrong Nash equilibrium condition as $k$ \emph{linear} inequalities and the original problem \eqref{eq:spoa_bound} in an LP,
3) The bound is proven tight by constructing an example using a solution $\theta^\star$.
The number of decision variables and constraints in \eqref{opt:primal} is $\mathcal{O}(n^3)$, making the problem solvable for larger numbers of agents.

To bring greater context to this result, we consider a specific class of resource allocation problems.
Consider the class of covering problems in which $w(x) = \indicator[x>0]$, i.e., a resource's value is fully contributed to the welfare if at least one agent utilizes it.
This particular setting has been well studied in the literature.
When $k=1$, it is known that the Nash equilibrium approximates the optimal welfare within a factor of $1/2$~\cite{Vetta2002}.
When $k=n$, it is clear that the only \kstrong Nash equilibrium are those that maximize the system welfare, thus $\spoa_k=1$.
For coalitions of size $1 < k < n$, the \kstrong price of anarchy can be found via \cref{thm:quant}.

In \cref{fig:spoa}, we show the tight bound on the \kstrong price of anarchy in covering problems with $n=20$, for each $k \in [n]$.
The endpoints of this curve are an efficiency of 1/2 when agents are fully decentralized and an efficiency of 1 with full collaboration.
\cref{fig:spoa} shows how we can bridge this gap with inter-agent collaboration; making clear that collaborative decision-making has the opportunity to improve system performance.
These gains in system performance do come at the cost of added complexity in the form of group decision-making, which will be the subject of future work.
\cref{thm:quant} informs us that despite these communication and computation costs, opportunities for improvement do exist.
Additionally, it is possible to further improve the benefit of collaboration.
The decision-making process described in \cref{sec:model} has each agent (or group of agents) maximize the welfare function; in \cref{subsec:design} we consider designing the common-interest objective the agents maximize.

\begin{figure*}
\begin{align}
Q^\star = &\min_{\mu, \{u_\zeta \in \mathbb{R}_{\geq 0}^{n}\}_{\zeta \in [k]}} \quad && \mu\nonumber\\
&~~{\rm s.t.} && \hspace{-50pt}w(o\hs +\hs x)-\mu w(e\hs +\hs x) + \sum_{\zeta \in [k]}\hs  \left( \hs \frac{n!}{(n\hs -\hs \zeta)!} u_\zeta(e\hs +\hs x) -\hs  \hs \sum_{\substack{0\leq\alpha\leq \zeta\\ 0\leq\beta\leq \zeta-\alpha}} \hs {\zeta\choose \alpha} {\zeta-\alpha \choose \beta} e^{\underline{\alpha}}o^{\underline{\beta}}(n\hs-\hs e\hs -\hs o)^{\underline{\zeta\text{-}\alpha\text{-}\beta}} u_\zeta (e\hs +\hs x\hs +\hs \beta\hs -\hs \alpha)  \hs \right) \leq 0 \nonumber\\
& && \hspace{4.2in}\forall (e,x,o) \in \mathcal{I}\tag{Q}\label{opt:design}
\end{align}
\vspace{-8mm}
\end{figure*}

\subsection{Utility Design}\label{subsec:design}
In \cref{sec:model}, we described a solution to a distributed decision-making problem of assigning agents (and groups of agents) to maximize the system objective.
Though this may appear like a reasonable approach, it need not be optimal.
As a method of further improving the benefit of inter-agent collaboration, we will consider designing the agents utility functions as a method to alter the emergent equilibria.
Consider a common-interest \textit{utility function} \hbox{$U:\mathcal{A} \rightarrow \mathbb{R}$.}
Following the coalitional best response dynamics, if a group $\Gamma \subseteq N$ is selected to update their action, they now update according to
\begin{equation}
a_\Gamma \in \argmax_{a_\Gamma^\prime \in \mathcal{A}_\Gamma} U(a_\Gamma^\prime,a_{-\Gamma}).
\end{equation}
As the multi-agent decision model remains a common-interest game (now with payoff $U(a)$), the existence and convergence results of \cref{prop:exist} hold.
With the new objective for the agents, the set of \kstrong Nash equilibria is altered to $\KSNE(G,U)$.
Changing the utility function of the agents has no impact on the original system objective $W$; our goal is to select the function $U$ in a way such that the new equilibria offer a better approximation of the optimal system objective $W$.
To quantify how well this is done, we extend the definition of \kstrong price of anarchy to depend on the utility rule, i.e.,
\begin{equation}\label{eq:spoa_design}
    \spoa_k(G,W,U) = \frac{\min_{\KSNash{a} \in \KSNE(G,U)} W(\KSNash{a})}{\max_{\Opt{a} \in \mathcal{A}} W(\Opt{a}) }.
\end{equation}

Again, we focus our attention to the class of resource allocation problems as described in \cref{subsec:define}.
As these problems are large and combinatorial, rather than designing a global utility function for each joint action, we will consider local utility rules of the form $u:\{0,\ldots,n\} \rightarrow \mathbb{R}$.
With this, the agents' common-interest objective becomes,
\begin{equation}
    U(a) = \sum_{r \in \mathcal{R}} v_r u(|a|_r).
\end{equation}
Recall that this does not change the overall system objective \eqref{eq:ra_welf}; $u$ simply alters agents' decision-making.
We set $u(0)=0$ else the \kstrong price of anarchy will be zero.

As before, we are interested in the performance guaranteed by a collaborative decision-making system.
As such, we will consider the efficiency guarantee across the class of instances $\gee_n$ with a welfare function $w$.
When using utility rule $u$, the \kstrong price of anarchy will be denoted
$\spoa_k(\gee_n,w,u) = \min_{G \in \gee_n} \spoa_k(G,w,u)$.
We are particularly interested in designing $u$ to provide us the best guarantee; let
\begin{equation}\label{eq:spoa_star_design}
    \spoa_k^\star(\gee_n,w) = \max_{u:[n] \rightarrow \mathbb{R}} \spoa_k(\gee_n,w,u),
\end{equation}
denote the optimal \kstrong price of anarchy, or the approximation ratio of a \kstrong Nash equilibrium under the optimal utility design.

In \cref{prop:design}, we provide an upper bound on the efficiency an optimally designed utility rule can provide.
\begin{proposition}\label{prop:design}
For the class of resource allocation problems $\gee_n$ with welfare function $w$, when agents maximize the optimal utility design objective $u^\star$,
\begin{equation}
\spoa_k^\star(\gee_n,w) \leq 1/Q^\star(n,w,k),
\end{equation}
where $Q^\star(n,w,k)$ is the value of the linear program \eqref{opt:design}.
\end{proposition}
The proof appears in the appendix.
The decision variables $\{u_\zeta\}_{\zeta \in [k]}$ of \eqref{opt:design} are the utility function for groups of size $\zeta \in [k]$.
If $u_\zeta = u_{\zeta^\prime}$ for all $\zeta,\zeta^\prime \in [k]$, the value of \eqref{opt:design} is the exact solution to \eqref{eq:spoa_star_design}.
However, this need not be the case in general, hence \cref{prop:design} is an upper bound.
The authors conjecture that the class of homogeneous local utility rules can attain a $\spoa_k^\star$ of $1/Q^\star(n,w,k)$, however proving this is the subject of ongoing work.



To bring greater context to this, we again consider the case of covering problems with objective $w(x) = \indicator[x>0]$.
When $k=n$, it is clear the optimal utility design is to let the group of all agents maximize the common-interest objective $w$, leading to an $n$-strong price of anarchy of $\spoa_n^\star(\gee_n,\indicator) = 1$.
When $k=1$, the authors of \cite{Gairing2009} show that the optimal utility rule gives price of anarchy $\spoa_1^\star(\gee_n,\indicator) = 1-\frac{1}{\frac{1}{(n-1)(n-1)!}+\sum_{j=0}^{n-1}\frac{1}{j!}}$ which approximately equals $0.632$ when $n=20$; this increase from $1/2$ when agents maximize the system welfare motivates the fact that designing agents' utilities can provide performance improvements.
In \cref{fig:spoa}, we use \cref{prop:design} to show the possible benefit of utility design for groups of size $1 \leq k \leq n$.
By comparing the naive approach of setting agents to maximize the system objective and the results of \cref{prop:design}, it appears that significant benefit can occur from designing how groups make decisions.
Future work will seek to quantify the optimal utility rule in closed form.

\section*{Conclusion}
In this work, we provide a tractable linear program whose solution is a tight bound on the \kstrong price of anarchy in resource allocation games.
Additionally, we exploit this program to provide an upper bound on the largest attainable performance when agents' utility function can be designed.
Future work will study the added computational complexity of reaching these equilibria and deriving bounds in closed form when the number of agents is arbitrary but the size of coalitions is a constant fraction of the number of agents.

\bibliographystyle{IEEEtran}
\bibliography{../../../My_Library}

\appendix

\iftrue
\noindent\emph{Proof of \cref{prop:exist}:}
To show existence, we can simply observe that $\Opt{a} \in \argmax_{a \in \mathcal{A}} W(a)$ is a \kstrong Nash equilibrium for any $k \in [n]$.
Because $W(\Opt{a})\geq W(a^\prime)$ for all $a^\prime \in \mathcal {A}$, the global optimal satisfies
$$W(\Opt{a}) \geq W(a_{\Gamma}^\prime,\Opt{a}_{-\Gamma}),~\forall a_\Gamma^\prime \in \mathcal{A}_\Gamma,~\Gamma \in \mathcal{C}_{[k]}.$$

For the deterministic coalition best response dynamics, let $a \in \mathcal{A}$ be the current action.
Let a `round' be defined by each subset $\Gamma \in \mathcal{C}_{[k]}$ being given the opportunity to revise their action $a_\Gamma \in \mathcal{A}_\Gamma$.
By definition, any action revision strictly increases the system welfare.
If a round passes and no subset revises, the current state $a$ is a \kstrong Nash equilibrium.
Because there are finitely many joint-actions $a \in \mathcal{A}$, the deterministic coalition best response dynamics will reach a state where no agent revises in at most $|\mathcal{A}|$ rounds.

For the asynchronous coalition best response, a subset $\Gamma \in \mathcal{C}_{[k]}$ is selected uniformly at random to revise their action.
A group $\Gamma$ revises their action to one of strictly higher payoff if one exists.
Consider the resulting Markov chain $\mathcal{M}$ with states $\mathcal{A}$.
Any state $a \in \mathcal{A} \setminus \KSNE$ has an outgoing edge with positive probability as there exists some group $\Gamma \in \mathcal{C}_{[k]}$ that is selected with probability $1/|\mathcal{C}_{[k]}|$ which would revise their action.
Any state $a \in \KSNE$ has no outgoing edges with positive probability as no group $\Gamma\in \mathcal{C}_{[k]}$ can revise their action to strictly increase the welfare.
Finally, there are no cycles (excluding self loops) in $\mathcal{M}$, as every outgoing edge is directed from a joint action of lower welfare to one of higher welfare.
As such, the set $\KSNE$ is absorbing and $\mathbb{P}[\lim_{t \rightarrow \infty} a(t) \in \KSNE] = 1$.
\hfill\qed
\fi

\noindent\emph{Proof of \cref{thm:quant}:} The proof can be outlined by three parts: first, the problem of finding $\spoa_k(\gee_n,w)$ is transformed and relaxed, second, a parameterization is introduced to turn the relaxed problem into a linear program, finally, an example is constructed via the solution to the linear program to show the linear program provides a tight bound.

\noindent\emph{1) Relaxing the problem:} Quantifying $\spoa_k(\gee_n,w)$ can be expressed as taking the minimum \kstrong price of anarchy over all games in $\gee_n$, i.e.,
\begin{mini}|s|
{\scriptstyle G \in \gee_n}{\frac{\min_{\KSNash{a} \in \KSNE(G)} W(\KSNash{a})}{\max_{\Opt{a} \in \mathcal{A}} W(\Opt{a})}}
{\tag{P1}\label{opt:tproof1}}{}
\end{mini}
To make this problem more approachable, we introduce several transformations and relaxations.
First, rather than searching over the entire set of game $\gee_n$, we search over the set of games $\hat{\gee}_n$, in which each agent has exactly two actions.
This reduction of the search-space can be done without loss of generality, i.e., $\spoa_k(\gee_n,w) = \spoa_k(\hat{\gee}_n,w)$.
Trivially, $\hat{\gee}_n \subset \gee_n$.
Further consider any game $G \in \gee_n$; if for every player each of their actions is removed except their action in the optimal allocation $\Opt{a}_i$ and their action in their worst \kstrong Nash equilibrium $\KSNash{a}_i$, the new problem will maintain the same \kstrong price of anarchy, but will now exist in $\hat{\gee}_n$.
With this reduction, we will denote each player's action set as $\hat{\mathcal{A}}_i = \{\Opt{a}_i,\KSNash{a}_i\}$.
Second, we normalize each resource value $v_r$ such that the equilibrium welfare is one.
This too can be done without loss of generality by scaling each resource identically thus not altering the $\spoa$ ratio.
Third, we invert the objective and consider the maximization of $W(\Opt{a})/W(\KSNash{a})$.
Finally, we sum over each of the $k$-coalition equilibrium constraints.
For each $\zeta \in [k]$, rather than satisfying each inequality in \eqref{eq:ksnash_def}, sum over every permutation\footnote{The sum over permutations rather than combinations is redundant compared to \cref{def:ksnash}, but the feasible set is still generalized.} of the $\zeta$ out of $n$ players, denoted $\mathcal{P}_\zeta$.
Applying these reductions to \eqref{opt:tproof1} gives,
\begin{maxi}|s|
{\scriptstyle G \in \hat{\gee}_n}{W(\Opt{a})}
{\tag{P2}\label{opt:tproof2}}{}
\addConstraint{\hs\hs\frac{n!}{(n\hs-\hs\zeta)!}W(\KSNash{a}) \geq \sum_{\Gamma \in \mathcal{P}_\zeta} W(\Opt{a}_\Gamma,\KSNash{a}_{-\Gamma}),~\forall \zeta \in [k]}
\addConstraint{W(\KSNash{a}) = 1}{}
\end{maxi}
\eqref{opt:tproof2} provides a lower bound on $\spoa_k(\gee_n,w)$ as the feasible set was expanded.
Later, we will show that the bound is tight by constructing an example that realizes it.

\noindent\emph{2) Parameterization:} To each resource $r \in \mathcal{R}$, we assign a label $(e_r,x_r,o_r)$, where
\begin{align*}
    e_r &= \lvert\{ i \in N \mid r \in \KSNash{a}_i \setminus \Opt{a}_i\}\rvert \\
    x_r &= \lvert\{ i \in N \mid r \in \KSNash{a}_i \cap \Opt{a}_i\}\rvert \\
    o_r &= \lvert\{ i \in N \mid r \in \Opt{a}_i \setminus \KSNash{a}_i\}\rvert.
\end{align*}
This is to say, $e_r$ denotes the number of agents utilizing resource $r$ in only their equilibrium action, $o_r$, is the number that uses resource $r$ in only their optimal action, and $x_r$ is the number that use $r$ in both their equilibrium and optimal actions.
In the set of games $\gee_n$, let $\mathcal{I} = \{(e,x,o)\in \mathbb{N}_{\geq 0}^3 \mid 1 \leq e+x+o \leq n\}$ denote the set of possible labels, and
$\theta(e,x,o) := \sum_{r \in \mathcal{R}_{(e,x,o)}} v_r,$
where $\mathcal{R}_{(e,x,o)} = \{r \in \mathcal{R} \mid e_r = e, x_r = x, o_r=o\}$ denotes the set of resources with label $(e,x,o)$.
The parameter $\theta \in \mathbb{R}_{\geq 0}^{|\mathcal{I}|}$ is a vector with elements for each label.

We will now express the terms in \eqref{opt:tproof2} using this parameterization.
Because $W(a) = \sum_{r \in \mathcal{R}} v_r w(|a|_r)$ depends only on the number of agents utilizing a resource, we can represent $|\Opt{a}|_r = e_r + x_r$ and write the optimal welfare as
\begin{align*}
    W(\Opt{a}) &= \sum_{r \in \mathcal{R}}v_r w(o_r+x_r)\\
    &=\sum_{(e,x,o) \in \mathcal{I}} \Bigg( \sum_{r \in \mathcal{R}_{e,x,o}} v_r \Bigg)w(o+x)\\
    &= \sum_{e,x,o}\theta(e,x,o)w(o+x).
\end{align*}
When not stated, the sum over $(e,x,o)$ is implied to be for each label in $\mathcal{I}$.
Similar steps can be followed to show $W(\KSNash{a}) = \sum_{e,x,o} \theta(e,x,o)w(e+x)$.

Finally, we will represent $\sum_{\Gamma \in \mathcal{P}_\zeta} W(\Opt{a}_\Gamma,\KSNash{a}_{-\Gamma})$ with the introduced parameterization.
First, we group all resources that are associated with the same label.
Consider one such label $(e,x,o)$.
Let $z \in \{\mathbf{e},\mathbf{o},\mathbf{nx}\}^\zeta$ denote a registry of how each player in a coalition of size $\zeta$ may use a resource, i.e., if agent $i \in \Gamma$ in the coalition uses the resource in only their equilibrium action, then $z_i = \mathbf{e}$; similarly if player $i$ only uses $r$ in their optimal action, $z_i = \mathbf{o}$, or if in neither or both of these actions $z_i = \mathbf{nx}$. The value of the elements is not relevant, it simply will be used to record how agents use a resource.
A coalitions interaction with a resource can be described solely by $z$.
Let $z_\mathbf{e} := |\{ z_i \in z \mid z_i = \mathbf{e}\}|$ be the number of players in a coalition that interact with the resource as type $\mathbf{e}$.
Define $z_\mathbf{o}$ similarly.
Now, we will group coalitions that interact with the resource in the same way.
When a coalition $\Gamma$ interacts with a resource $r$ with label $(e,x,o)$ as described by $z$, then the utilization of $r$ satisfies $|\Opt{a}_\Gamma,\KSNash{a}_{-\Gamma}|_r = e+x-z_\mathbf{e}+z_\mathbf{o}$.
There are $n!/(n-\zeta)!$ permutations of players of size exactly $\zeta$; of those, 
$$B_{(e,x,o)}^{\zeta,n}(z_{\textbf{e}},z_{\textbf{o}}):={\zeta \choose z_\mathbf{e}} {\zeta-z_\mathbf{e} \choose z_\mathbf{o}} e^{\underline{z_\mathbf{e}}}o^{\underline{z_\mathbf{o}}}(n-e-o)^{\underline{\zeta-z_\mathbf{e}-z_\mathbf{o}}},$$
have $z_\mathbf{e}$ players using $r$ as type $\mathbf{e}$ and $z_\mathbf{o}$ using $r$ as type $\mathbf{o}$.
With this, observe that
\begin{align*}
\sum_{\Gamma \in \mathcal{P}_\zeta} & W(\Opt{a}_\Gamma,\KSNash{a}_{-\Gamma})\\
 &= \sum_{\Gamma\in\mathcal{P}_\zeta} \sum_{e,x,o} \sum_{r \in \mathcal{R}_{(e,x,o)}} \hspace{-2mm} v_r w(|\Opt{a}_\Gamma,\KSNash{a}_{-\Gamma}|_r)\\
&= \sum_{e,x,o} \sum_{r \in \mathcal{R}_{(e,x,o)}}\hspace{-2mm} v_r \sum_{\Gamma\in\mathcal{P}_\zeta} w(|\Opt{a}_\Gamma,\KSNash{a}_{-\Gamma}|_r)\\
&=  \sum_{e,x,o} \sum_{r \in \mathcal{R}_{(e,x,o)}}\hspace{-2mm} v_r \sum_{\substack{0 \leq \alpha \leq \zeta\\ 0\leq \beta \leq \zeta-\alpha}} \hspace{-2mm} B_{(e,x,o)}^{\zeta,n}(\alpha,\beta)w(e \hspace{-1mm} + \hspace{-1mm} x \hspace{-1mm} + \hspace{-1mm} \beta \hspace{-1mm} - \hspace{-1mm} \alpha)\\
&= \sum_{e,x,o} \theta(e,x,o) \sum_{\substack{0 \leq \alpha \leq \zeta\\ 0\leq \beta \leq \zeta-\alpha}} \hspace{-2mm} B_{(e,x,o)}^{\zeta,n}(\alpha,\beta)w(e \hspace{-1mm} + \hspace{-1mm} x \hspace{-1mm} + \hspace{-1mm} \beta \hspace{-1mm} - \hspace{-1mm} \alpha)
\end{align*}
Note that if $e+x+\beta-\alpha < 0$ (where $w()$ is not defined), then $B_{(e,x,o)}^{\zeta,n}(\alpha,\beta) = 0$ and one can disregard these terms in the sum.
Substituting terms using the described parameterization gives \eqref{opt:primal}.
Note, that any two problems that are encoded by the same vector $\theta$ have the same \kstrong price of anarchy.
As such, $\theta$ becomes the lone decision variable.
The problem of solving for $\spoa_k(\gee_n,w)$ has been transformed to a linear program in $\theta$; however, the problem was relaxed and thus \eqref{opt:primal} gives a lower bound.
Next, we will show that this bound is tight by constructing an example that realizes the value of the linear program.

\begin{figure}
    \centering
    \vspace{1mm}
    \includegraphics[width=0.4825\textwidth]{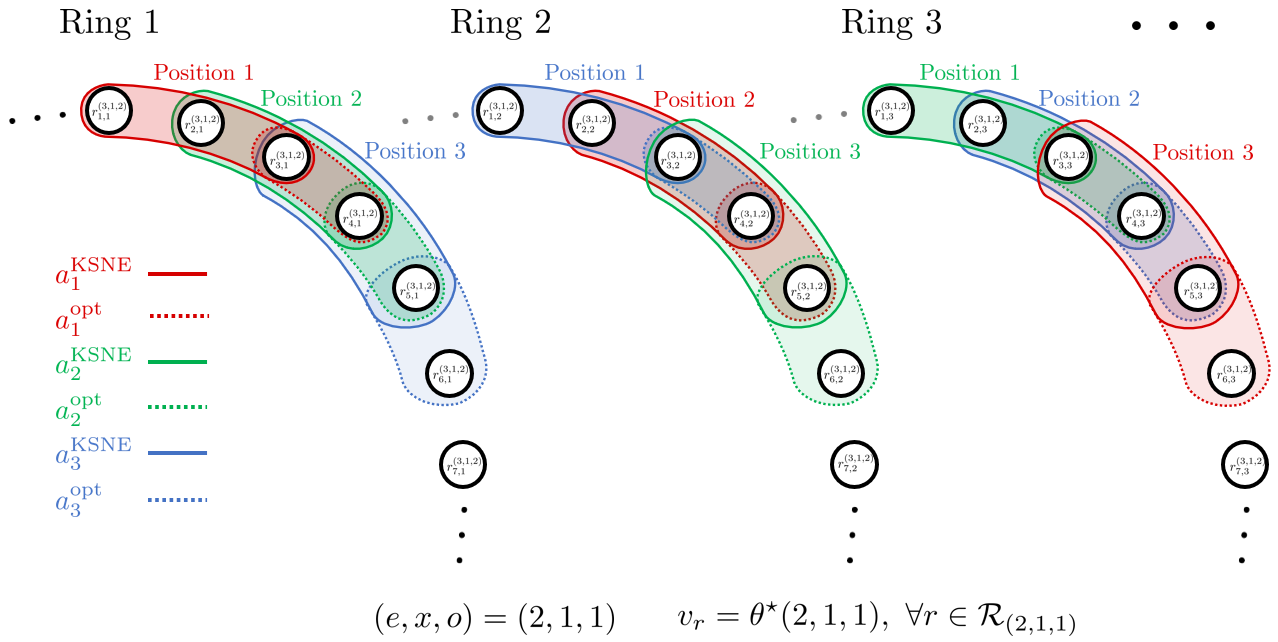}
    \caption{Game construction for worst-case \kstrong price of anarchy. Three of the $n$ players' action sets are shown (color coded in red, green, and blue respectively) on three of $n!$ rings for the label $(e,x,o)=(2,1,1)$. A ring has $n$ positions, one for each player. For a label $(e,x,o)$ we generate $n!$ rings for all the orderings of players over positions. This is repeated for each label. Players still only have two actions, but each action covers resources from each ring. The value of a resource is equal to the value of $\theta^\star$, the solution to \eqref{opt:primal}, for the label with which it is associated.}
    \label{fig:rings}
    \vspace{-6mm}
\end{figure}

\noindent\emph{3) Constructing an example:} Consider the following resource allocation problem: for a label $(e,x,o) \in \mathcal{I}$, consider $n$ resources labeled $r_{1,1}^{(e,x,o)}$ to $r_{n,1}^{(e,x,o)}$.
For these resources, we define $n$ positions that the players can take: if player $i$ is in position 1, then they use resources $\{r_{1,1}^{(e,x,o)},\ldots,r_{e+x,1}^{(e,x,o)}\}$ in their action $\KSNash{a}_i$ and resources $\{r_{e+1,1}^{(e,x,o)},\ldots,r_{e+x+o,1}^{(e,x,o)} \}$ in their action $\Opt{a}_i$. 
The remaining positions take the same form, but incremented around the ring, i.e.,
if player $j$ is in position 2, then they use resources $\{r_{2,1}^{(e,x,o)},\ldots,r_{e+x+1,1}^{(e,x,o)}\}$ in their action $\KSNash{a}_j$ and resources $\{r_{e+2,1}^{(e,x,o)},\ldots,r_{e+x+o+1,1}^{(e,x,o)} \}$ in their action $\Opt{a}_j$, and so on for the remaining positions up to $n$
Each player $i \in N$ is assigned a position on this ring; now, repeat this ring structure $n!$ times for each possible ordering of players over the positions.
For the label $(e,x,o)=(2,1,1)$, this is depicted for the first three rings and three players in \cref{fig:rings}.
Repeat this process for each label $(e,x,o) \in \mathcal{I}$.
Finally, let the value of each resource with label $(e,x,o)$ be $v_{r^{(e,x,o)}_{\cdot,\cdot}} = \theta^\star(e,x,o)$, where $\theta^\star$ is a solution to \eqref{opt:primal}.
Each player still has two actions, but they select resources from each ring in each action, changing actions differentiates which resources in the rings they select.
We will now show 2 things about this game: 1) $\KSNash{a}$ is a \kstrong Nash equilibrium, 2) $\Opt{a}$ has welfare $P^\star(n,w,k)$.

For the first part, consider the welfare in the state $\KSNash{a}$: each resource in a ring with label $(e,x,o)$ has $e+x$ players utilizing it in $\KSNash{a}$.
There are $n n!$ such resources for each label, as such, 
\begin{equation}\label{eq:W_ksne_welf}
W(\KSNash{a}) = \sum_{e,x,o}n n!\theta^\star(e,x,o)w(e+x).
\end{equation}
Now consider any group $\Gamma \in \mathcal{C}_{[k]}$, and let $\zeta = |\Gamma|$.
Let $c = \{c_1,\ldots,c_\zeta\}$ denote a position of the group on some ring, i.e., $c_i \in [n]$ is the position of player $i \in \Gamma$.
Consider a label $(e,x,o)$; there are $(n-\zeta)!$ rings in which group $\Gamma$ is in configuration $c$.
Consider all the resources with the first index ($r_{1,\cdot}^{(e,x,o)}$) in each ring with label $(e,x,o)$, if each player $i \in \Gamma$ deviates their action to $\Opt{a}_i$, then of these resource $B_{(e,x,o)}^{\zeta,n}(\alpha,\beta)$ have $\alpha$ players from $\Gamma$ use them as a type $\textbf{e}$ player and $\beta$ players from $\Gamma$ use them as a type $\textbf{o}$ player.
The welfare contributed by one of these resources is $\theta^\star(e,x,o)w(e+x-\alpha+\beta)$.
Summing over all values of $0 \leq \alpha \leq \zeta$ and $0 \leq \beta \leq \zeta-\alpha$ gives the welfare of all the resources with first index, i.e. $(n-\zeta)!\theta^\star(e,x,o)\sum_{0 \leq \alpha \leq \zeta} \sum_{0\leq \beta \leq k-\alpha}B_{(e,x,o)}^{\zeta,n}(\alpha,\beta)$.
Due to the symmetry of the rings, the welfare of the resource in each position is the same.
Summing over all labels gives
\begin{align}
&W(\Opt{a}_\Gamma,\KSNash{a}_{-\Gamma})\label{eq:W_gamma_dev}\\
 &= n (n\hspace{-1mm}-\hspace{-1mm}\zeta)! \sum_{e,x,o}\theta^\star(e,x,o) \hspace{-3mm}\sum_{\substack{0 \leq \alpha \leq \zeta\\ 0\leq \beta \leq \zeta-\alpha}} \hspace{-4mm}B_{(e,x,o)}^{\zeta,n}(\alpha,\beta)w(e \hspace{-1mm} + \hspace{-1mm} x \hspace{-1mm} + \hspace{-1mm} \beta \hspace{-1mm} - \hspace{-1mm} \alpha)\nonumber
\end{align}
Comparing \eqref{eq:W_ksne_welf} and \eqref{eq:W_gamma_dev} gives one of the $k$ inequality constraints in \eqref{opt:primal}.
Because $\theta^\star$ is a solution to \eqref{opt:primal}, and thus feasible, the constraint is satisfied, and the selected group $\Gamma$ would not generate greater welfare from deviating.
Again, by the symmetry of the problem and the fact the constraints exist for $\zeta \in [k]$, any group $\Gamma \in \mathcal{C}_{[k]}$ will have no positive deviations.
Further, if one considered only part of a group deviating, it is equivalent to only considering the sub-group fully deviating, and again can be shown to satisfy the inequalities in \eqref{opt:primal}.

Finally, consider the allocation $\Opt{a}$.
In any ring with label $(e,x,o)$, each resource has $o+x$ players utilizing it.
As such,
\begin{equation}\label{eq:W_opt_welf}
	W(\Opt{a}) = \sum_{e,x,o} n n!\theta^\star(e,x,o)w(o+x).
\end{equation}
Comparing \eqref{eq:W_ksne_welf} and \eqref{eq:W_opt_welf} produces
\begin{align*}
\frac{W(\KSNash{a})}{W(\Opt{a})} &= \frac{\sum_{e,x,o}n n!\theta^\star(e,x,o)w(e+x)}{\sum_{e,x,o} n n!\theta^\star(e,x,o)w(o+x)}\\
 &= \frac{1}{\sum_{e,x,o}\theta^\star(e,x,o)w(o+x)} = \frac{1}{P^\star(n,w,k)},
\end{align*}
where $\sum_{e,x,o}\theta^\star(e,x,o)w(e+x) =1 $ by $\theta^\star$ being feasible in \eqref{opt:primal}, and $P^\star(n,w,k)$ is the solution to \eqref{opt:primal}.
The \kstrong price of anarchy in this problem is at most $1/P^\star(n,w,k)$.
\hfill\qed

\noindent\emph{Proof of \cref{prop:design}:}
First, we express \eqref{opt:primal} more concisely:
\begin{maxi}|s|
{\scriptstyle \theta \in \mathbb{R}^{|\mathcal{I}|}}{b^\top \theta\nonumber}
{}{}
\addConstraint{c_\zeta^\top \theta }{\geq 0,~\forall \zeta \in [k]}{\quad\quad(\lambda_\zeta)}
\addConstraint{d^\top \theta - 1 }{= 0}{\quad\quad(\mu)}
\addConstraint{\theta}{\geq 0}{\quad\quad(\rho)}
\end{maxi}
where $\lambda \geq 0$, $\mu$, and $\rho \geq 0$ are the associated dual variables.
Note that the primal is feasible and bounded; the feasibility holds from the existence of \kstrong equilibria from \cref{prop:exist}, and the boundedness holds from $w(y)>0$ for all $y>0$ having non-zero price of anarchy~\cite{Paccagnan2018a}.
The Lagrangian function is defined as $\mathcal{L}(\theta,\lambda,\mu,\rho) = b^\top \theta + (\sum_{\zeta \in [k]} \lambda_\zeta c_\zeta^\top\theta) - \mu (d^\top \theta -1) + \rho^\top \theta$.
Let $g(\lambda,\mu,\rho) = \sup_{\theta \in \mathbb{R}^{|\mathcal{I}|}} \mathcal{L}(\lambda,\mu,\rho)$ serve as an upper bound to \eqref{opt:primal}.
The dual program is derived by minimizing $g(\lambda,\mu,\rho)$; note that this value is only unbounded above unless $b^\top + \sum_{\zeta \in [k]}\lambda_\zeta c_\zeta^\top - \mu d^\top + \rho^\top = 0$.
Substituting this into the objective, and removing the free variable $\rho$ so that the equality constraint becomes an inequality, the dual problem becomes
\begin{mini}|s|
{\scriptstyle \mu,\{\lambda_\zeta \in \mathbb{R}_{\geq 0}\}_{\zeta \in [k]}}{\mu}
{\tag{Q1}\label{opt:design1}}{}
\addConstraint{b^\top - \mu d^\top + \sum_{\zeta \in [k]} \lambda_\zeta c_\zeta }{\leq 0}
\addConstraint{}
\end{mini}
From strong duality, \eqref{opt:design1} provides the same value as \eqref{opt:primal}.

Finally, we consider that in each term $c_\zeta$ (associated with the equilibrium constraint), we replace $w(\cdot)$ with $\hat{u}_\zeta(\cdot)$.
Selecting $\hat{u}_\zeta = w$ for each $\zeta \in [k]$ would give the same program.
Further, let $u_\zeta(j) := \lambda_\zeta \hat{u}(j)$ for all $j \in [n]$ and replace this in the constraint of \eqref{opt:design1}.
The resulting constraint is now linear in each $u_\zeta$.
Minimizing the result of \eqref{opt:design1} (equivalent to maximizing SPoA as it is the reciprocal of the solution) over $\{u_\zeta\}_{\zeta \in [k]}$ gives \eqref{opt:design}.
This program optimizes over all $\{u_\zeta \in \mathbb{R}_{\geq 0}^{n}\}_{\zeta \in [k]}$ or all utility rules for each coalition size $\zeta \in [k]$, where the utility rule for different sized coalitions may differ.
Because the feasible set of $\{u_\zeta \in \mathbb{R}_{\geq 0}^{n}\}_{\zeta \in [k]}$ contains $w$,
\eqref{opt:design} provides a lower bound $Q^\star(n,w,k)$, and thus an \hbox{upper bound on $\spoa_k^\star(\gee_n,w)$.
\hfill\qed}

\end{document}